\documentclass[a4paper]{article}

\usepackage{comment}
\usepackage[left=2.25cm, right=2.25cm, top=2cm, bottom=2cm, marginparwidth=1.7cm, marginparsep=2mm]{geometry}
\usepackage[utf8]{inputenc}
\usepackage[T1]{fontenc}
\usepackage{cmbright}
\usepackage{graphicx}
\usepackage[table, dvipsnames]{xcolor}
\usepackage{amsfonts}
\usepackage{amsmath}
\usepackage{amsthm}
\usepackage{mathtools, nccmath}
\usepackage{url}
\usepackage{booktabs}
\usepackage{tabularx}
\usepackage{nicefrac}
\usepackage{microtype}
\usepackage[hidelinks]{hyperref}
\hypersetup{
    colorlinks=true,
    linkcolor=blue,
    urlcolor=blue,
    citecolor=blue,
    anchorcolor=blue}
\usepackage[inline]{enumitem}
\usepackage{float}
\usepackage{wrapfig}
\usepackage[font=footnotesize, labelfont=bf]{caption}
\usepackage{subcaption}
\usepackage{adjustbox}
\usepackage{tikz}
\usetikzlibrary{shapes.geometric}
\usepackage{changepage}
\usepackage{etoc}
\usepackage{ragged2e}
\usepackage[framemethod=TikZ]{mdframed}
\usepackage[super, comma, numbers]{natbib}
\usepackage[symbol]{footmisc}
\usepackage{multicol}
\usepackage{orcidlink}
\definecolor{orcidlogocol}{HTML}{000000}

\newcounter{theo}[section] 
\renewcommand{\thetheo}{\arabic{theo}}
\newenvironment{theo}[2][]{%
\small
\refstepcounter{theo}%
\ifstrempty{#1}%
{\mdfsetup{%
frametitle={%
\tikz[baseline=(current bounding box.east),outer sep=0pt]
\node[anchor=east, rectangle, fill=blue!20]
{\strut Box~\thetheo};}}
}%
{\mdfsetup{%
frametitle={%
\tikz[baseline=(current bounding box.east), outer sep=0pt]
\node[anchor=east, rectangle, fill=blue!20]
{\strut Box~\thetheo:~#1};}}%
}%
\mdfsetup{innertopmargin=5pt, innerbottommargin=10pt, linecolor=blue!20, linewidth=2pt, frametitleaboveskip=\dimexpr-\ht\strutbox\relax}
\begin{mdframed}[]\relax%
\label{#2}
\footnotesize}
{\end{mdframed}}

\definecolor{darkblue}{HTML}{1A254B}
\definecolor{lightblue}{HTML}{A7BED3}
\definecolor{blue}{HTML}{114083}
\definecolor{green}{HTML}{81B5AE}
\definecolor{pink}{HTML}{F2545B}
\definecolor{red}{HTML}{A4243B}
\definecolor{lightgray}{HTML}{C3BABA}
\definecolor{darkgray}{HTML}{9A8F97}

\title{\vspace{-30pt} \bf How to Build the Virtual Cell with Artificial Intelligence: \\ Priorities and Opportunities}
\date{}

\author{%
\small Charlotte Bunne$^{\orcidlink{0000-0003-1431-103X}\,1,2,3,4,}$\footnotemark[1]\,, 
\small Yusuf Roohani$^{\orcidlink{0000-0002-9577-5014}\,1,3,5}$\footnotemark[1]\,,
\small Yanay Rosen$^{\orcidlink{0000-0003-1129-9645}\,1,3,}$\footnotemark[1]\,,
\small Ankit Gupta$^{\orcidlink{0000-0002-9961-1041}\,3,6}$,
\small Xikun Zhang$^{\orcidlink{0000-0002-8346-8594}\,1,3,7}$,
\small Marcel Roed$^{\orcidlink{0000-0002-1369-8933}\,1,3}$, \\
\small Theo Alexandrov$^{8,9}$,
\small Mohammed AlQuraishi$^{10}$,
\small Patricia Brennan$^{3}$,
\small Daniel B. Burkhardt$^{11}$,
\small Andrea Califano$^{10,12,13}$, \\
\small Jonah Cool$^{3}$,
\small Abby F. Dernburg$^{14}$,
\small Kirsty Ewing$^{3}$,
\small Emily B. Fox$^{1,15,16}$,
\small Matthias Haury$^{17}$,
\small Amy E. Herr$^{16,18}$, \\
\small Eric Horvitz$^{19}$,
\small Patrick D. Hsu$^{5,18,20}$,
\small Viren Jain$^{21}$,
\small Gregory R. Johnson$^{22}$,
\small Thomas Kalil$^{23}$,
\small David R. Kelley$^{24}$, \\
\small Shana O. Kelley$^{25,26}$,
\small Anna Kreshuk$^{27}$,
\small Tim Mitchison$^{28}$,
\small Stephani Otte$^{17}$,
\small Jay Shendure$^{29,30,31,32}$, \\
\small Nicholas J. Sofroniew$^{33}$,
\small Fabian Theis$^{34,35,36}$,
\small Christina V. Theodoris$^{37,38}$,
\small Srigokul Upadhyayula$^{14,16,39}$, \\
\small Marc Valer$^{3}$,
\small Bo Wang$^{40,41}$,
\small Eric Xing$^{42,43}$,
\small Serena Yeung-Levy$^{1,44}$,
\small Marinka Zitnik$^{45,46,47}$, \\
\small Theofanis Karaletsos$^{\orcidlink{0000-0002-0296-3092}\,3,}$\footnotemark[3]\,,
\small Aviv Regev$^{\orcidlink{0000-0003-3293-3158}\,2,}$\footnotemark[3]\,,
\small Emma Lundberg$^{\orcidlink{0000-0001-7034-0850}\,3,6,7,48,}$\footnotemark[3]\,,
\small Jure Leskovec$^{\orcidlink{0000-0002-5411-923X}\,1,3,}$\footnotemark[3]\,,
\small Stephen R. Quake$^{\orcidlink{0000-0002-1613-0809}\,3,7,49,}$\footnotemark[3] \\[2ex]}

\begin{document}
\maketitle

\vspace{-30pt}
{\scriptsize \noindent
\footnotemark[1] These authors contributed equally. \\
\footnotemark[3] Correspondence to: \href{mailto:tkaraletsos@chanzuckerberg.com}{Theofanis Karaletsos}, \href{mailto:regev.aviv@gene.com}{Aviv Regev}, \href{mailto:emmalu@stanford.edu}{Emma Lundberg}, \href{mailto:jure@cs.stanford.edu}{Jure Leskovec}, and \href{mailto:mailto:quake@stanford.edu}{Stephen R. Quake}. \\

\noindent $^{1}$ Department of Computer Science, Stanford University, Stanford, CA, USA,
$^{2}$ Genentech, South San Francisco, CA, USA, 
$^{3}$ Chan Zuckerberg Initiative, Redwood City, CA, USA,
$^{4}$ School of Computer and Communication Sciences and School of Life Sciences, EPFL, Lausanne, Switzerland,
$^{5}$ Arc Institute, Palo Alto, CA, USA,
$^{6}$ KTH Royal Institute of Technology, Science for Life Laboratory, Department of Protein Science, Stockholm, Sweden,  
$^{7}$ Department of Bioengineering, Stanford University, Stanford, CA, USA, 
$^{8}$ Department of Pharmacology, University of California, San Diego, CA, USA,  
$^{9}$ Department of Bioengineering, University of California, San Diego, CA, USA,
$^{10}$ Department of Systems Biology, Columbia University, New York, NY, USA, 
$^{11}$ Cellarity, Somerville, MA, USA, 
$^{12}$ Vagelos College of Physicians and Surgeons, Columbia University Irving Medical Center, New York, NY, USA,  
$^{13}$ Chan Zuckerberg Biohub New York, NY, USA,
$^{14}$ Department of Molecular and Cell Biology, University of California, Berkeley, Berkeley, CA, USA, 
$^{15}$ Department of Statistics, Stanford University, Stanford, CA, USA, 
$^{16}$ Chan Zuckerberg Biohub San Francisco, CA, USA,  
$^{17}$ Chan Zuckerberg Institute for Advanced Biological Imaging, Redwood City, CA, USA, 
$^{18}$ Department of Bioengineering, University of California, Berkeley, CA, USA,
$^{19}$ Microsoft Research, Redmond, WA, USA,   
$^{20}$ Center for Computational Biology, University of California, Berkeley, Berkeley, CA, USA, 
$^{21}$ Google Research, Mountain View, CA, USA, 
$^{22}$ NewLimit, San Francisco, CA, USA, 
$^{23}$ Schmidt Futures, USA, 
$^{24}$ Calico Life Sciences LLC, San Francisco, CA, USA,
$^{25}$ Chan Zuckerberg Biohub Chicago, IL, USA, 
$^{26}$ Northwestern University, Evanston, IL, USA,  
$^{27}$ Cell Biology and Biophysics Unit, European Molecular Biology Laboratory, Heidelberg, Germany, 
$^{28}$ Department of Systems Biology, Harvard Medical School, Boston, MA, USA,  
$^{29}$ Department of Genome Sciences, University of Washington, Seattle, WA, USA, 
$^{30}$ Brotman Baty Institute for Precision Medicine, Seattle, WA, USA,  
$^{31}$ Seattle Hub for Synthetic Biology, Seattle, WA, USA,
$^{32}$ Howard Hughes Medical Institute, Seattle, WA, USA, 
$^{33}$ EvolutionaryScale, PBC, USA, 
$^{34}$ Institute of Computational Biology, Helmholtz Center Munich, Munich, Germany,  
$^{35}$ School of Computing, Information and Technology, Technical University of Munich, Munich, Germany,  
$^{36}$ TUM School of Life Sciences Weihenstephan, Technical University of Munich, Munich, Germany,  
$^{37}$ Gladstone Institute of Cardiovascular Disease, 
Gladstone Institute of Data Science and Biotechnology, San Francisco, CA, USA,
$^{38}$ Department of Pediatrics, University of California, San Francisco, CA, USA,  
$^{39}$ Molecular Biophysics and Integrated Bioimaging Division, Lawrence Berkeley National Laboratory, Berkeley, CA, USA,  
$^{40}$ Department of Computer Science, University of Toronto, Toronto, Ontario, Canada,  
$^{41}$ Vector Institute, Toronto, Ontario, Canada,  
$^{42}$ Carnegie Mellon University, School of Computer Science, Pittsburgh, PA, USA,  
$^{43}$ Mohamed Bin Zayed University of Artificial Intelligence, Abu Dhabi, United Arab Emirates,  
$^{44}$ Department of Biomedical Data Science, Stanford University, Stanford, CA, USA,  
$^{45}$ Department of Biomedical Informatics, Harvard Medical School, Boston, MA, USA,  
$^{46}$ Kempner Institute for the Study of Natural and Artificial Intelligence, Harvard University, Cambridge, MA, USA,  
$^{47}$ Broad Institute of MIT and Harvard, Cambridge, MA, USA,
$^{48}$ Department of Pathology, Stanford University, Stanford, CA, USA,  
$^{49}$ Department of Applied Physics, Stanford University, Stanford, CA, USA. \\

}

\begin{multicols}{2}

\begingroup
\section*{Abstract}
\vspace{-9pt}
\small
\looseness -1 The cell is arguably the most fundamental unit of life and is central to understanding biology. Accurate modeling of cells is important for this understanding  as well as for determining the root causes of disease. Recent advances in artificial intelligence (AI), combined with the ability to generate large-scale experimental data, present novel opportunities to model cells. Here we propose a vision of leveraging advances in AI to construct virtual cells, high-fidelity simulations of cells and cellular systems under different conditions that are directly learned from biological data across measurements and scales. We discuss desired capabilities of such AI Virtual Cells, including generating universal representations of biological entities across scales, and facilitating interpretable \emph{in silico} experiments to predict and understand their behavior using Virtual Instruments. 
We further address the challenges, opportunities and requirements to realize this vision including data needs, evaluation strategies, and community standards and engagement to ensure biological accuracy and broad utility. We envision a future where AI Virtual Cells help identify new drug targets, predict cellular responses to perturbations, as well as scale hypothesis exploration. With open science collaborations across the biomedical ecosystem that includes academia, philanthropy, and the biopharma and AI industries, a comprehensive predictive understanding of cell mechanisms and interactions has come into reach.
\par\endgroup
\bigskip\noindent

\section*{Main}

The cell, the fundamental unit of life, is a wondrously intricate entity with properties and behaviors that challenge the limits of physical and computational modeling. Every cell is a dynamic and adaptive system in which complex behavior emerges from a myriad of molecular interactions. Some aspects are remarkably robust to perturbations, such as the elimination of genes or their replacement with homologs from different species. Other aspects are sensitive to  even seemingly minor disruptions, such as a point mutation or an external factor that tip cells into dysfunction and disease.

\begin{figure*}[ht]
    \centering
    \includegraphics[width=.9\textwidth]{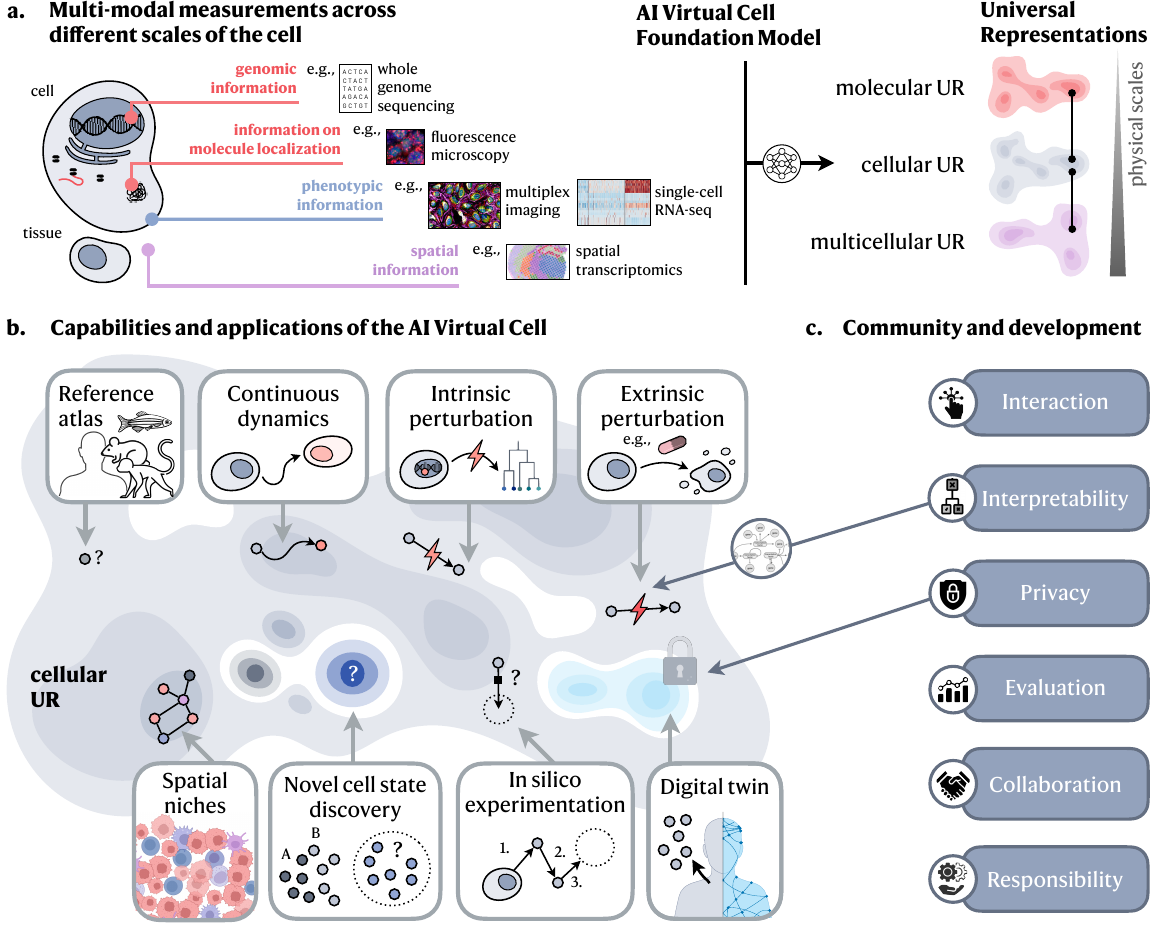}
    \caption{\looseness -1 \textbf{Capabilities of the AI Virtual Cell.}
    \textbf{a.} The AI Virtual Cell provides a Universal Representation of a cell state that can be obtained across species and conditions, and generated from different data modalities across scales (molecular, cellular, multicellular). \textbf{b.} The AI Virtual Cell possesses capabilities to represent and predict cell biology. 
    This universality allows the representation to act as a reference that can generalize to previously unobserved cell states, providing guidance for future data generation. Since the representation is shared across modalities, it also remains invariant to the specific data type used to generate it, serving as a virtual representation for unified analysis across modalities. 
    The AI Virtual Cell also allows modeling the dynamics of cells as they transition between different states, whether naturally due to processes such as differentiation or due to genetic variation or artificially through engineered perturbations. Thus, the AI Virtual Cell enables \emph{in silico} experimentation that would otherwise be cost-prohibitive or impossible in a lab.
    \textbf{c.} The utility of the AI Virtual Cell depends on its interactions with humans at different levels. At the individual scientist level, it must be accessible through open licenses and the democratization of computing resources. Interpretability can be established through intermediary layers such as language models that allow the virtual cell to communicate its results effectively. At the scientific community level, evaluating the AI Virtual Cell should focus on core capabilities that move beyond narrow benchmarks. Community development will be crucial for ongoing improvements to the virtual cell that remain accessible. At the societal level, the AI Virtual Cell must ensure the privacy of its contents to protect sensitive data.
    }
    \label{fig:capabilities}
\end{figure*}

To understand a cell's function, scientists have attempted to construct \emph{virtual cell} models to simulate, predict and steer cell behavior \citep{slepchenko2003quantitative, johnson2023building, marx_2023, goldberg2018emerging, georgouli2023multi, marucci_2020}. 
Building on this vision, we use the term \emph{virtual cell} to define a computational model that simulates the biological functions and interactions of a cell. 
Existing cell models are often rule-based and combine assumptions about the underlying biological mechanisms with parameters fit from observational data. They generally rely on explicitly-defined mathematical or computational approaches, such as differential equations \citep{alon2019introduction, lauffenburger1996receptors, karr2012whole, mangan2003structure, sachs2005causal}, stochastic simulations \citep{zopf2013cell, eling2019challenges, to2010noise} or agent-based models \citep{hellweger2016advancing, gorochowski2016agent}. 
They vary in complexity and cover different defined aspects of cell biology such as transcription \citep{woodcock_2013} and translation \citep{thiele_2009}, cytoskeletal driven cell behavior \citep{odell_2008, popov_2016}, biochemical networks \citep{burke_2020} or metabolic flux \citep{li_2023, fang_2020}. The first whole-cell model was developed in 2012, representing all 482 genes and molecular functions known for an organism; the bacteria \emph{Mycobacterium genitalium} \citep{karr2012whole}. Since this pioneering work, genome-wide models have been developed to represent other bacterial organisms, including \emph{Escherichia coli} \citep{fang_2020, karr2012whole, stevens_2023, maritan_2022, ahnhorst_2022}. 

Despite their widespread use in modeling biological systems, approaches to date fall short of capturing many aspects of the operations of both bacteria and more complex systems, such as human cells: 
(1) Multi-scale modeling: Cells operate on multiple scales across both time and space, from atomic to molecular to cellular and histological, with functional properties emerging through nonlinear transformation from one scale to another.
(2) Diverse processes with massive numbers of interacting components: Cellular function encompasses numerous interacting processes, such as gene regulation, metabolic pathways and signal transduction. Each process involves a multitude  of biomolecular species, in diverse and dynamic configurations and states.
(3) Nonlinear dynamics: Most cellular processes are highly nonlinear, such that small changes in inputs can lead to complex changes in outputs. 
Thus, despite progress in modeling specific cellular processes, these factors collectively pose a substantial roadblock to the construction of a virtual cell.

Two exciting revolutions in science and technology---in AI and in 'omics---now enable the construction of cell models learned directly from data. These parallel revolutions provide an unprecedented opportunity for an ambitious vision of an \emph{AI Virtual Cell} (AIVC), a multi-scale, multi-modal, large neural network-based model that can represent and simulate the behavior of molecules, cells and tissues across diverse states (Fig.~
\ref{fig:capabilities}).

\looseness -1 Experimentally, the exponential increase in the throughput of measurement technologies has led to the collection of large and growing reference datasets within and across different cell and tissue systems  \citep{collins1998new, venter2001sequence, regev_2017, czi2023cz}, with data doubling every six months for the past several years \citep{heimberg_2023}, along with the ability to couple these measurements with systematic perturbations \citep{dixit2016perturb, srivatsan2020massively, feldman2022pooled, bock2022high}. 

\looseness -1 Computationally, concurrent advances in AI have enhanced our ability to learn patterns and processes directly from data without needing explicit rules or human annotation \citep{vaswani_2017, rombach_2021, brown_2020}. Such modeling paradigms have been used successfully in the biomolecular realm, for example to predict three-dimensional molecular structure from sequence \citep{jumper_2021, baek2021accurate, lin2023evolutionary, townshend2021geometric} and interactions between different molecular components \citep{gomes2017atomic, alipanahi2015predicting, gainza2020deciphering, cunningham2020biophysical, torng2019high, corso_2022}. Recent modeling methodologies in AI provide representation and inference tools that satisfy the trifecta of being predictive, generative and queryable, which are key utilities for biological research and understanding. By building on these properties, we argue that we now have the methods to develop a fully data-driven neural network-based representation of an AI Virtual Cell that can accelerate research in biomedicine by enabling fast-paced \emph{in silico} studies, as well as powerful bridges between computational methods and confirmatory wet-lab experimentation (Fig.~\ref{fig:capabilities}).


The creation of an AIVC will enable a new era of high-fidelity simulation in biology, in which cancer biologists model how specific mutations transition cells from healthy to malignant; developmental biologists forecast how developmental lineages evolve in response to perturbation in specific progenitor cells; microbiologists predict the effects of viral infection on  not just the infected cell but also its  host organism. AIVCs will empower experimentalists and theorists alike, by transforming the means by which hypotheses are generated and prioritized, and allowing biologists to span a dramatically expanded scope, better fitting the enormous scales of biology. 
Although the cellular models may not always directly identify mechanistic relationships, they can be viewed as tools for effectively narrowing the search space for mechanistic hypotheses, thereby accelerating the discovery of underlying factors behind cellular function. 
 
This Perspective article is based on extensive community discussions, including a workshop hosted by the Chan Zuckerberg Initiative, and aims to ignite the formation of a collaborative research agenda for a large-scale, long-term initiative with a roadmap for developing, implementing, and deploying AI Virtual Cells. We describe a vision catalyzed by emerging advances in AI in cell biology and their application to constructing virtual representations of cells. We lay out priorities and opportunities across data generation, AI models, benchmarking, interpretation and  ensuring biological veracity and safety (Box~\ref{box:challenges}). By encouraging interdisciplinary collaborations in open science---spanning academia, philanthropy, and the biopharma and AI industries---we posit that a comprehensive understanding of cellular mechanisms is within reach. AI Virtual Cells have the potential to revolutionize the scientific process, lead to the understanding of novel biological principles and augment human intelligence to underpin future breakthroughs in programmable biology, drug discovery and personalized medicine (Box~\ref{box:vignettes}).

\section*{AI Virtual Cells}

Our view of an AI Virtual Cell is a learned simulator of cells and cellular systems under varying conditions and changing contexts, such as differentiation states, perturbations, disease states, stochastic fluctuations, and environmental conditions (Fig.~\ref{fig:capabilities}). In this context, a virtual cell should integrate broad knowledge across cell biology. Virtual cells must work across biological scales, over time, and across data modalities, and should help reveal the programming language of cellular systems and provide an interface to use it for engineering purposes.

\looseness -1 In particular, an AI Virtual Cell needs to have capabilities that allows researchers to: 
(1) Create a universal representation of biological states across species, modalities, datasets and contexts, including cell types, developmental stages and external conditions;
(2) Predict cellular function, behavior and dynamics, as well as uncovering the underlying mechanisms;
(3) Perform \emph{in silico} experiments to generate and test new scientific hypotheses and guide data collection to efficiently expand the virtual cell's abilities.


Next, we elaborate on these key capabilities and discuss approaches for how to achieve them.



\subsection*{Universal representations}

An AIVC would map biological data to universal representational spaces (Fig.~\ref{fig:capabilities}a), facilitating insights into shared states and serving as a comprehensive reference. These universal representations should integrate across three physical scales: molecular, cellular and multicellular, and accommodate contributions from any relevant modality and context  (Fig.~\ref{fig:capabilities}a). This integration will allow researchers to complement new data with existing information within the AIVC, leveraging its extensive biological knowledge to bridge gaps between different data. Such a comparison with  prior data would provide a comprehensive context for every analysis.

Importantly, the multilevel representation should generalize to new states not present within the data used to train the AIVC. Such an emergent capability would unlock discoveries about biological states that have not been directly observed, or might not even occur in nature. For instance, the AIVC's exposure to similar instances during training, like inflammatory states in macrophages, might enable it to predict previously unknown inflammatory states in microglia. 
Additionally, the AIVC should be able to predict novel states resulting from interventions (or, equivalently, interventions needed to achieve a novel specified state) offering a range of downstream applications in cell engineering and synthetic biology.


\subsection*{Predicting cell behavior and understanding mechanisms}

\looseness -1 A defining function of an AIVC will be its ability to model cellular responses and dynamics. By training on a wide range of snapshots, time-resolved, non-interventional and interventional datasets collected across contexts and scales, the AIVC can develop an understanding of the molecular, cellular, and tissue dynamics that occur under natural or engineered signals. These signals include external and internal stresses or other factors like chemical (e.g., small molecules) or genetic (engineered or natural) perturbations and their combinations. An AIVC should be able to predict responses to perturbations that have not been previously tested in the lab, while also accounting for the specific features of the cellular context within which the perturbation is being tested. 

The AIVC should also have the capability to simulate the temporal evolution of alterations in cell states in response to both intrinsic or extrinsic factors, along with the resulting multicellular spatial arrangements. By modeling the transient nature of the overall cell state and the continuous flux in cellular conditions, the AIVC could uncover previously unstudied trajectories in diverse dynamic processes like development, maintenance of homeostasis, pathogenesis and disease progression.


Another critical challenge is understanding the molecular mechanisms underpinning observed phenotypes and trajectories. The AIVC could propose potential causal factors behind phenotypes by simulating the effects of different interventions.
Through its multi-scale design, the AIVC should be able to extrapolate the basis of cellular function across scales, and 
link intracellular processes to phenotypes at the cell and tissue level. Thus, the AIVC opens new avenues for investigating mechanisms linked to diverse phenotypes and behaviors.

Although uncovering a phenotype's causal factors may not always be feasible through computation alone, the AIVC has the potential to reduce the space of possible hypotheses. Through simulating the effects of different interventions, the AIVC could propose potential causal factors behind phenotypes with corresponding degrees of uncertainty, allowing scientists to validate claims experimentally.

\subsection*{\emph{In silico} experimentation and guiding data generation}

For real world utility, a defining function of an AIVC will be its ability to guide data generation and experiment design. An AIVC should be queryable with computational twins of today’s laboratory experiments, here called \textit{Virtual Instuments}. Virtual experiments could, for example, simulate experiments in a cell type that is challenging to cultivate \textit{in vitro}, or simulate expensive readouts from low-cost measurements, such as single cell transcriptomes from label free imaging \citep{kudo2023highly}.  Virtual experiments could also be used to screen a vast number of possible perturbagens, at a scale that would be impossible in the lab. Such capabilities are invaluable when considering the exponentially larger search space of combinatorial perturbations involving more than one perturbagen \citep{roohani_2023, bunne2023learning, lotfollahi2023predicting, bunne2022supervised, bereket2024modelling}.


AIVCs will usher in a new paradigm of how computational systems are probed during the design of new biological experiments. In this framework, an AIVC would not only design experiments to validate specific scientific hypotheses, but also to enhance its own capabilities.  Equipped with the ability to assign confidence values to its predictions, an AIVC could enable interactive querying to guide  experimentalists to the most efficient path for generating additional data for fine-tuned improvement in low-confidence areas.  Extended to an active and iterative lab-in-the-loop process, we envision  efficient and focused expansion of the AIVC's 
performance. Ultimately, the AIVC might even be able to identify key gaps in its own understanding of biology and propose the most efficient paths to bridge them \citep{huang_2023, hie2020leveraging, roohani_2024, cleary2020necessity}.


\section*{Building the AIVC}


We envision an AI Virtual Cell as a comprehensive AI framework composed of several interconnected foundation models that represent dynamic biological systems at increasingly complex levels of organization---from molecules to cells, tissues, and beyond. 
Our approach has two main components: (1) a universal multi-modal multi-scale biological state representation and (2) a set of virtual instruments, which are neural networks that manipulate or decode these representations. While there may be other approaches to building an AIVC, we believe this approach would provide a scaffold that can be scaled in a collaborative and open way.

We use the term \textit{Universal Representation} (UR) to refer to an embedding produced by a multi-modal AIVC foundation model. An embedding is a learned numerical representation of data in a continuous vector space. The AIVC transforms high dimensional multi-scale multi-modal biological data into embeddings that retain meaningful relationships and patterns.

The AIVC can capture cell biology at three distinct physical scales by representing (1) molecules and their structures found within individual cells, (2) individual cells, as spatial collections of those interacting molecules and structures, and (3) how individual cells interact with one another and the non-cellular environment in a tissue. Each of these scales is represented by a distinct UR, building on abstractions generated by the previous layer, thus linking the different scales. 

In the context of UR, \textit{Virtual Instruments} (VIs) are neural networks that take URs as input and produce a desired output. We describe two types of VIs:
Decoder Virtual Instruments (or Decoders) that take a UR as an input and produce human-understandable output, for example, a cell type label or a synthetic microscope image; and,
Manipulator Virtual Instruments (or Manipulators) which take a UR as an input and produce another UR as an output, for example, that of an altered cell state after perturbation. Since these instruments will operate over the same representations, they can be shared and reused across different use cases, experiments and datasets. Thus, we envision that any scientist will be able to build a Virtual Instrument on top of a Universal Representation and share it with the community. The building of VIs that closely resemble real instruments, like a microscope, has the potential to seed the development of instrument specific lab-in-the-loop systems. 


\begin{figure*}[t]
    \centering
    \includegraphics[width=\textwidth]{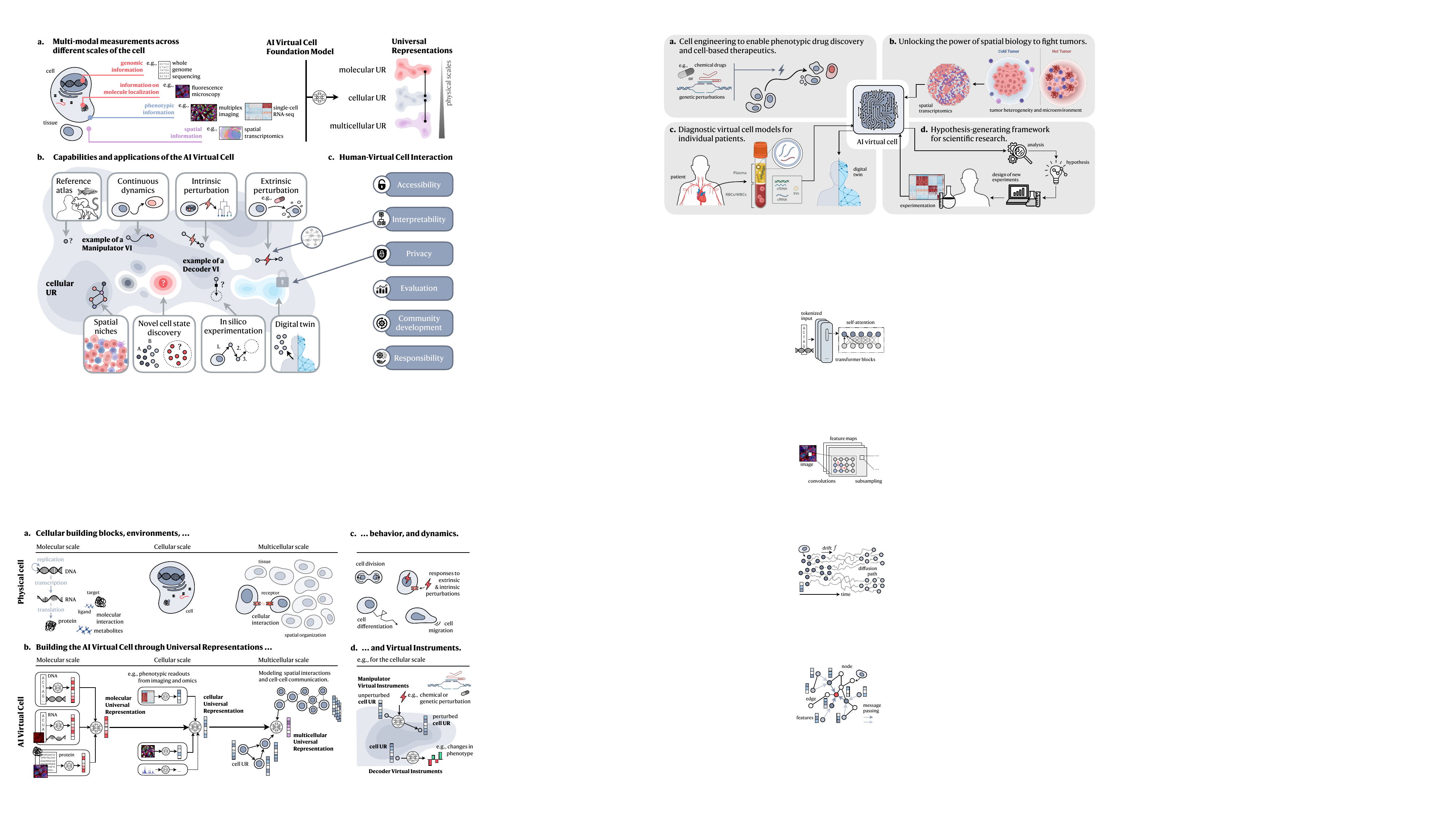}
    \caption{\looseness -1 \textbf{Overview of the AI Virtual Cell.}
    \textbf{a.} Similar to biological cells, \textbf{b.} the AI Virtual Cell models cell biology across different physical scales, including molecular, cellular, and multicellular. Along the physical dimension, the first scale models the state and interactions of individual molecules, such as those of the central dogma, as well as additional molecules like metabolites. Molecules can be represented as sequences or atomic structures. The next scale represents cells as collections of these molecules. For example, such cells contain a genetic sequence, RNA transcripts and some quantities of proteins. Molecules within cells have specific locations that may be related to their function. The final scale models the interactions between cells, how they communicate and form complex tissues. Each scale relies on Universal Representations that are learned from multi-modal data and are integrating URs from the previous scale.  \textbf{c.} To capture the behavior and dynamics of physical cells, its components, or collections, \textbf{d.} the AI Virtual Cell comprises Virtual Instruments. On the cellular scale, for example, Manipulator VIs simulate how cell states change as cells divide, migrate, develop from progenitor states, or respond to perturbations through learned transitions in the URs. Decoder VIs allow to decode the cell UR, e.g., to understand phenotypic properties.
    }
    \label{fig:virtualcell}
\end{figure*}

\subsection*{Building universal representation across physical scales}

The AIVC would be a multi-scale foundation model that learns distinct representations of biological entities at each physical scale (Fig.~\ref{fig:virtualcell}c). These representations can be aggregated together and transformed to produce representations at the next higher physical scale. This recurring architectural motif can be applied from the level of individual molecules to the scale of entire tissues and organs granting the model consistency across biological scales (Fig.~\ref{fig:virtualcell}a). Each representation applies universally to that specific class of biological entities. This abstraction allows the virtual cell to seamlessly evolve and incorporate new data, from new modalities or from out-of-distribution sources, since it can all integrate within this general framework.


In the following, we discuss design principles and data that could be used to construct each physical scale of the AIVC bottom-up. While many existing machine learning architectures could be applied directly to the task of learning functional representations of cellular components (Box~\ref{box:architectures}), we additionally suggest the incorporation of biological inductive biases into the design of these representations, and further modeling innovations should drive the refinement and success of these models.

\looseness -1  \paragraph{Molecular scale.} 
The first layer of the virtual cell represents individual molecular species (Fig.~\ref{fig:virtualcell}a,c). While there are many different classes of molecules present in a cell, a starting point for the AIVC will be to model the 
three types of molecules of the central dogma: DNA, RNA and proteins. These can all be represented as sequences of characters; nucleotides or amino acids \citep{ji_2021, rives2021biological, brandes_2022, celaj_2023, dallatorre_2023, nguyen_2023, nguyen_2024, hayes_2024, ruffolo_2024}. Such sequence data are particularly well-suited for AI methods originally developed for natural language processing, like large language models (LLMs) (Box~\ref{box:architectures}). Given the high-throughput measurement capabilities for DNA, RNA and protein sequences, there are substantial and growing amounts of training data available. This abundance of data, combined with simple objective functions (such as predicting masked letters in a sequence), provides the key ingredients for effectively training models to generate an initial molecular UR. Furthermore, a biological language model could be trained on all three modalities simultaneously, thus maximizing interoperability and training corpus size. Despite its inherent compatibility with transformers, specific considerations around masking and attention mechanisms must be addressed when applying these models to biological sequence data as opposed to natural language. 

While language modeling approaches have been extensively studied for these core molecules and have proven successful for some of their chemical modifications \citep{peng_2024_ptm} and various other molecules such as glycans, lipids and metabolites \citep{dai_2021, yu2024lipidbert}, they may struggle with other  molecular constituents of the cell. Such modeling difficulties might be exacerbated for data that is difficult to fit into a sequence, or very small molecules. Given that the primary building blocks of these entities are atoms, a neural network trained to model molecules at the atomic level \citep{jumper_2021, krishna_2024} could be a more general choice for this layer. However, models with atomic resolution introduce a substantial  computational burden and might be constrained by limited availability of training data.

\paragraph{Cellular scale.}

The next level of abstraction models individual cell states (Fig.~\ref{fig:virtualcell}a,c). As cellular function is underpinned by the molecular interactions and signaling networks formed in a cell, a cellular UR can be built using representations of molecular and other (e.g., imaging) features describing the organization and abundance of molecular components. The key step here would be to integrate learned representations of molecules with their quantities, and appropriately abstracted locations and timestamps, to create a unified representation of the cell~\citep{rosen2024saturn, rosen_2023, chen_2024_genept}.

\looseness -1 Data for the cellular UR consist of measurements mapped to a single cell level such as measurements of the transcriptome (scRNA-seq), chromatin accessibility (scATAC-seq), chromatin modification  transcription factor binding, and proteome \citep{mahdessian2021spatiotemporal}. 
 Imaging technologies measure cell morphology at subcellular resolution, often together with molecular information \citep{chandrasekaran2024three, carlson2023genome, feldman_2022}. For example, fluorescence confocal microscopy can help resolve the subcellular location of the human proteome \citep{thul2017subcellular}. Live-cell imaging \citep{mcdole2018toto} enables the study of proteins in living cells using time-lapse microscopy. Cryo-electron microscopy determines biomolecular structures at near-atomic resolution \citep{nogales2024bridging, bauda2024ultrastructure}. Super-resolution microscopy offers deeper insights into molecular processes through single-molecule imaging in living systems \citep{lelek2021single, mockl2020super, deniz2008single}. Complementing imaging approaches, mass spectrometry and proximity-dependent labeling can unveil protein-protein associations and provide deeper insights into cell structure and signaling network rewiring \citep{cesnik2024mapping, qin2021multi}. 

From the model architecture perspective, vision transformers \citep{dosovitskiy_2020} or models leveraging convolutional neural networks (CNNs) \citep{fukushima_1980, lecun_1995} are widely applicable to biological images to model across multiple imaging channels capturing different biological features \citep{moshkov2024learning, bao2024channel, le2022analysis, kraus2024masked}, while being robust to distribution shift and batch variability \citep{bao2023contextual}. Autoencoders and transformers have been successfully applied for learning representations for sequence-based data \citep{lopez2018deep, eraslan2019deep, rosen_2023, cui2024scgpt, theodoris2023transfer}. Using AI algorithms to integrate different data modalities collected with sequencing and imaging technologies creates a multi-view model of the cell that can be both dynamic and predictive \citep{comiter2023inference, kobayashi2021raman2rna, ryu2024cross}.


As the AIVC model grows in complexity, it is crucial to also model cellular organelles and membraneless compartments \citep{saar2024protein} as units that play specific roles within the cell. Robustly capturing the functions of these  units is vital to ensure accurate predictions, mechanistic interpretability and model generalizability. 

Given their prevalence, the cellular UR will initially rely on transcriptomics measurements, while imaging modalities will be key for continued modeling of cellular spatial organization and dynamics.

\paragraph{Multicellular scale.}

At the third layer of abstraction, the AIVC models the organization of cells into a multicellular UR (Fig.~\ref{fig:virtualcell}a,c). This layer allows for the exploration of how cell-cell interactions, largely governed by spatial proximity, combine into tissues, organs and, ultimately, whole organisms. 
Multicellular interactions can be analyzed after tissue dissociation (such as in scRNA-seq) \citep{macosko2015highly} or \emph{in situ} in a 2D section or 3D volume, where the tissue structure is preserved. Building the AIVC will require integration across available modalities that provide spatial insights, i.e., both spatial molecular profiling as well as non-molecular tissue imaging data.

There are multiple methods to profile the spatial location of RNA \citep{staahl2016visualization}, 
and proteins \citep{lundberg_2019} in cells, along with various imaging methods for select molecular species (e.g., immunohistochemistry), or with stains for tissue strucutre alone (e.g., haematoxylin and eosin (H\&E)). Spatial molecular biology is currently a very active area of research and method development. While publicly available data are still limited, we foresee a rapid development in this domain providing multi-omic 2D and 3D datasets. A more generalized data generation effort together with open frameworks for spatial data \citep{marconato2024spatialdata} could greatly accelerate modeling at the multicellular scale.


The relative organization of cells within a 2D tissue section and 3D tissue volume can be represented using a graph or point cloud. The multicellular UR can be derived from such data using graph-learning techniques such as graph neural networks (GNNs) \citep{scarselli_2009} and equivariant neural networks (ENNs) \citep{satorras2021n}. For image-based data, convolutional neural networks or vision transformers can be applied (Box~\ref{box:architectures}). 

\subsection*{Predicting cell behavior and understanding mechanism}

Virtual instruments are the ``tools'' that operate on UR embeddings and perform various functions and tasks. By altering universal representations of molecules, cells and tissues, Manipulators can abstract complex dynamic processes (Fig.~\ref{fig:virtualcell}b) more simply as transitions between (distributions of) their representations (Fig.~\ref{fig:virtualcell}d). 
Similarly, Decoders can take an embedding of biological entities and predict one or more concrete properties, for example, physical structure, cell type/state, fitness, expression or drug response.

The design of a wide array of Manipulators provides us with an unprecedented set of tools for modeling cell behavior and dynamics: Generative AI approaches such as diffusion models \citep{somnath2023aligned} or autoregressive transformers \citep{katharopoulos2020transformers}, i.e., model architectures that capture heterogeneity and parameterize continuous time dynamics, can predict a future state or evolution of a cell or molecular state \citep{krishna_2024, abramson_2024} (Box~\ref{box:architectures}).
Using integrated data from time-lapse imaging \citep{mcdole2018toto}, gene expression profiles \citep{macosko2015highly, klein2015droplet, qiu2024single}, and other modalities, Manipulators can allow inferring the phenotypic progression from stem cell to differentiated cell, while capturing the influence of both genetic factors and environmental conditions ---through learned interpolations and extrapolations between multi-scale URs of different cell states.
Similarly, they allow predicting the effect of treatments on patients, given a virtual representation of a patient's molecular profile.

Furthermore, variations in cellular URs can be linked to corresponding changes in molecular states or their spatial localization, influenced by downstream factors like genetic variants or functional changes in proteins that are represented in a lower scale of the AIVC.
Leveraging the ability of Manipulators to model temporally-resolved molecular and cellular events, Decoders of the AIVC could potentially identify cellular components, molecular pathways, and their interactions that contribute to each prediction and process. As such, the multi-scale design of the AIVC may unveil mechanistic hypotheses of such processes.


\subsection*{\emph{In silico} experimentation and guiding data generation}

Manipulator Virtual Instruments operating in the UR space could further enable the exploration of a broad range of hypotheses through \emph{in silico} experiments that virtually perturb a cell model. This might be achieved by predicting changes in the URs following a perturbation prompt (Fig.~\ref{fig:virtualcell}d) \citep{roohani_2023, bunne2022supervised, bereket2024modelling, lotfollahi2023predicting}.

\looseness -1 The design of Manipulators that predict transitions in the UR upon an \emph{in silico} input can build on conditional generative models: Deep learning architectures like conditional deep generative models \citep{rombach_2021} allow \emph{generating} the desired UR based on the property or context of interest (Box~\ref{box:architectures}). Here, high-throughput perturbation screens ---based on RNA-seq \citep{ macosko2015highly, dixit2016perturb, norman2019exploring, srivatsan2020massively}, optical pooled screens (OPS) \citep{feldman2022pooled, kudo2023highly, lawson2017situ, lawson2021imaging}, or other technologies--- offer a rich resource through which the AIVC can be trained to predict these effects. By conditioning on specific perturbations---such as environmental changes, genetic mutations, or chemical treatments---the generative model might produce a new UR reflecting the predicted cellular response. This conditioning could be achieved through learned or pre-computed embeddings of the affected molecular targets. Chemical compounds, small molecules and metabolites could be embedded based on their chemical properties. Additionally, large language models trained on comprehensive scientific literature and biological databases, such as gene ontology or drug banks, could further provide a rich contextual background used for conditioning the generative model, e.g., through considering wide range of interactions and side effects.


\looseness -1 Virtual Instruments can be designed so that predictions are accompanied by estimates of model uncertainty \citep{mitra2019parameter, papamarkou2024position}. Under a Bayesian formulation of its predictive function, the predictions made for cell perturbation outcomes could include an uncertainty score, either implicitly via inference, deep kernels~\citep{d2021stein, ober2021promises}, or through explicit estimation of the full posterior over model parameters \citep{karaletsos2020hierarchical, kapoor2022uncertainty}.
Some practical approaches utilize model ensembles~\citep{lakshminarayanan2017simple} or conformal predictions~\citep{angelopoulos2021gentle}. 
By assigning specific confidence levels to its predictions, the AIVC can call methods for computing the expected value of additional data or approximations referred to in machine learning as active learning, to guide experimental data collection \citep{huang_2023, hie2020leveraging} for expanding its UR. Alternatively, methods for computing the expected value of information could be used to guide data generation with the goal of optimizing a desired biological property \citep{papamarkou2024position}. 
Lastly, through its ability to conduct \emph{in silico} experiments and suggest additional informative experiments, the AIVC could become an integrative part of lab-in-the-loop schemes. This allows not only for a seamless experimental validation of its predictions, but also a sequence of experiments, predictions and generations of hypotheses that gradually improve our systematic understanding of molecular circuits that drive biological functions.

\section*{Data needs and requirements}

 
A key consideration for the AIVC is which datasets and modalities must be collected to enable its effective construction. Unlike traditional experimental design, where data are generated to test specific scientific hypotheses, data collection for training the AIVC should be focused on ensuring the broad applicability and generalizability expected of the AIVC. To meet these ambitions, data would ideally span different domains and modalities, capture the heterogeneity and diversity of biological variability, and enable models to distinguish between technical (measurement) noise, stochastic biological variation and physiological differences.


\looseness -1 Data generation will require simultaneous exploration of temporal and physical scales, while allowing for system perturbations. Here, classical imaging technologies \citep{thul_2017, cho_2022, uhln_2015}, including  live-cell, and newer structural imaging technologies like cryo-electron tomography and soft X-Ray tomography \citep{berger_2023, nogales2024bridging, loconte_2023}, as well as novel spatial omics technologies \citep{moffitt_2022, vandereyken_2023, palla_2022}, offer opportunities to model biomolecules and functions across scales. Furthermore, biological processes span a vast range of timescales, from the fastest reactions happening in picoseconds, to a cell division happening in a day, tumor development occurring over years, and neurodegeneration over decades. The recent construction of universal cell atlases \citep{tabulasapiensconsortium_2022, uhln_2015} may serve as a powerful resource for modeling cellular behavior over longer time scales, such as tissue formation. 
New approaches will be needed to build comparable data sets which capture the behavior of cells on shorter times scales, e.g., through methods such as live-cell imaging. 
Besides molecular measurements, an important aspect of data collection will lie in the measurement of biophysical and biochemical cellular properties to provide boundaries of physical and chemical realism to the AIVC. 

 
Another important driver for the development of AI Virtual Cells will be multi-modal datasets. For example, datasets which bridge molecular and spatial scales, such as single cell transcriptomics data combined with histology to understand how cells interact and what molecular signatures underpin the formation of specialized spatial niches \citep{he_2020}. Further technological development is needed to collect multi-modal data that better captures the relationship between molecular signatures, cell behavior, cellular regulation and organization.


While a core interest of virtual cell modeling will focus on human datasets for the purpose of understanding disease and aiding the development of novel therapeutics, human datasets are limited in our ability to perform controlled experimentation and perturbations \emph{in vivo}. Here, the field of 3D tissue biology, including culture systems such as organoids, is emerging as a tool to study the complexities of tissue architecture and function \citep{bock_2021} in a 3D environment while allowing perturbations of the system. Another critical avenue to surpass this limitation will be to perform diverse, organism-wide profiles of species spanning evolutionary history, across perturbations and under various conditions \citep{tabulamurisconsortium_2018, thetabulamicrocebusconsortium_2021, lu_2023_aging_fly, li_2022_fly_cell_atlas, lange_2023_zebrahub}.

Finally, a key aspect of biological data generation will be the exploration of combinatorial spaces: biological spaces are commonly high dimensional, and enumerating their variants is intractable in general, e.g., when considering all possible variants of a genome. Even for combinations of a small number of entities, exemplified in the case of enumerating pairs or sets of perturbations \citep{cleary2020necessity, norman2019exploring}, experimental design becomes exceedingly challenging.
As combinatorial possibilities quickly expand well beyond what is practical experimentally, or even computationally, new methods for their exploration must be developed.
 
\paragraph{How much data is needed to build the AI Virtual Cell?}

The scale of raw biological data is undeniable, but so is the sheer nominal size of even one human cell system, making first principle estimates challenging. 

For instance, the Short Read Archive of biological sequence data holds over 14 petabytes of information \citep{katz_2022}, which is more than one thousand times larger than the dataset used to train ChatGPT \citep{achiam2023gpt}. Large parts of this data may be redundant, or have diminishing returns if used for training, and the scaling laws for models' performances must be investigated thoroughly.

\looseness -1 In addition to data size, data diversity is critical to ensure model quality \citep{ding_2024}. Data from humans and model organisms like mice and \emph{E. coli} are unequally represented in sequence and literature databases, which when used for training, encode strong species biases \citep{ding_2024}. Other biases, for example towards specific diseases or human ancestral populations could also reduce the impact of AIVC models \citep{liao_2023}. As virtual cell efforts mature, the dialogue between the scientists who develop models, those who generate experimental data and funding organizations must be further intensified.

\section*{Model evaluation}

A more important question for the development of AI  Virtual Cells may not be “How to build them?”, but rather “How to build trust in their competence and fidelity?”
To this end, a comprehensive and adaptable benchmarking framework will be needed.
Although various frameworks already exist for tackling specific biological questions (for example, protein structure prediction models \citep{abramson_2024} were developed in the context of the CASP evaluation framework), the AIVC will need to demonstrate generalizability across numerous biological contexts and downstream tasks. It must account for dynamic distributions that evolve due to environmental changes, infections, genetic variants and other such factors causing distribution shifts \citep{liu2021towards}. 


The evaluation of AI Virtual Cells should prioritize both generalizability as well as discovering new biology. Generalizability measures how well the model performs in unseen contexts such as novel cell types and genetic backgrounds. It can be evaluated through a cross-modal reconstruction task, such as predicting gene expression given the morphology of a previously unseen cell or the next image in a sequence of microscopy images of cell state. Assessing generalizability builds confidence in the AIVC's ability to capture core biological processes and understand how they vary across different contexts. Establishing such cross-modal benchmarks to link scales and modalities in cell biology is of imminent priority to the research community, as these tasks are both biologically useful and well-defined.

Ultimately, AIVC models should be judged on their ability to unlock new ways of understanding biology. Such an evaluation will ensure that model development is aligned with biological relevance. The most useful initial accomplishments will likely be to generate valuable testable hypotheses. For this purpose, validation datasets that are related to phenotypes that are  experimentally verifiable may be suitable, such as growth rate of cells, molecular profiles, disrupted protein-protein interactions, or transcription factor binding. 

As the capabilities of AI Virtual Cells improve, we must consider whether statistical measures of performance are adequate, or if  interpretability and biological causality would be core requirements.
 
\section*{Interpretability and interaction}
 
One of the hallmarks of scientific discovery in biology has been the creation of mechanistic models of a phenomenon under observation. When creating virtual cells, we may have to forgo our ability to build fully mechanistic models, in favor of learning interactions that will generalize from data and predict beyond the observations. However, it is still desirable to strive towards increased interpretability.

Every AIVC prediction could be substantiated with the corresponding  multi-scale interactions that determine resulting states, e.g., understanding how a cellular subsystem or protein complex is disrupted in a diseased tissue can aid development of therapeutic interventions \citep{zheng_2021, ma_2018, yu_2018}. The modular structure of the AIVC will enable researchers to pinpoint specific genes, proteins, or molecular processes involved in each predicted behavior. Patterns in the wiring of large models can also be leveraged to uncover combinatorial biological interactions, such as those between proteins, which can be projected to interpretable spaces without restricting the generality of the original model. While many capabilities of the AIVC rely on predictive tasks, generating mechanistic hypotheses could provide experimental routes to understand and explore the AIVC's predictions further, and will be vital for the adoption and use of AI Virtual Cells.

Ultimately, it will be of key interest to build an interactive layer for the AIVC that enables researchers of varying expertise to grasp and utilize its predictions effectively. AI Agents, built using large language models, could serve as virtual research assistants, providing an intuitive interface for non-experts \citep{gao_2024, ouyang2022training, roohani_2024}. Leveraging their extensive knowledge of scientific literature, these language models can offer deeper insights into the predictions made by the AIVC.


\section*{An open collaborative approach}

Creating an AIVC requires tremendous investment, diverse backgrounds and many iterations, and can only be advanced by a concerted open science effort.
As a scientific community, we must strive to ensure that both the development and usage of Virtual Cells are accessible and responsive to the entire scientific community. These efforts would greatly benefit from open data resources and data standards, a collaborative platform for cell modeling, and especially open benchmark datasets and common validation strategies to ensure their biological fidelity and real-world utility. Such a collaborative program could greatly accelerate progress across individual efforts and unify scientific research at a global scale, connecting myriad smaller-scale efforts.
 
\looseness -1 To achieve this, multiple key parameters need to be considered. First, we must ensure that AI Virtual Cells represent and benefit all of humanity, with open data that captures human ancestral and geographic diversity. Ensuring that such datasets reflect human diversity while safeguarding individuals' privacy is a principal challenge.
Second, as the size of AI Virtual Cell models increases, the cost of training, fine-tuning or using them as is will also grow. Investments in infrastructure and a platform for hosting these models will be critical to ensure accessibility and benefit to the broader scientific community. The platform should foster open and collaborative development of AI Virtual Cells, enabling active collaboration between biologists, clinicians and computer scientists. This platform should facilitate swift iterations between the lab and the modeling environment and offer opportunities to quickly test and benchmark new models.
Third, synergistic collaboration amongst stakeholders is needed across the biomedical ecosystem including philanthropy, academia, biopharma and the AI industry. Pre-competitive collaborations can greatly accelerate our collective progress towards creating AI Virtual Cells. Besides the synchronization with data generators and other modeling efforts, collaboration with regulatory authorities and bioethics experts are crucial for benchmarking and establishing new norms that will expedite the deployment of AI Virtual Cells, while complying with legal requirements and setting standards for ethical issues for responsible use of virtual cells.
 
This article is intended to serve as a primer for the formation of a collaborative research agenda and roadmap for a large-scale, long-term initiative for developing and implementing AI-powered Virtual Cells. If successful, such interactive AI Virtual Cell models, capable of simulating cellular biology, have the potential to fundamentally change how cell biology research is done. We foresee a future where AI Virtual Cell platforms function as open, interconnected hubs for collaborative development and broad deployment of cell models to researchers, but also as education hubs delivering training to researchers, as well as providing engagement activities for educators, patients and the public.
\section*{Outlook and reasons for optimism}

The genetics and genomics communities have created large reference datasets, such as the human genome project \citep{venter2001sequence}, HapMap \citep{gibbs2003international}, the Cancer Genome Atlas (TCGA) \citep{weinstein2013cancer}, ENCODE \citep{encodeprojectconsortium_2012}, the Genotype-Tissue Expression (GTEx) project \citep{gtexconsortium_2013}, the Human Protein Atlas (HPA) \citep{pontn_2008,thul2017subcellular}, the Human Cell Atlas (HCA) \citep{regev_2017} and a growing number of deeply phenotyped, population-scale biobank efforts \citep{downey_2008}. Thanks to these projects, massive reference data are now available to train machine learning models. While these efforts will continue to grow, they also catalyze a new, parallel effort: creating a virtual simulation of cell biology, a new process for scientific inquiry.

The result, the AI Virtual Cell has the potential to revolutionize the scientific process, leading to future breakthroughs in biomedical research, personalized medicine, drug discovery, cell engineering and programmable biology. 
Acting as a virtual laboratory, the AIVC could facilitate a seamless interface between data derived from \emph{in silico} experimentation and results from physical laboratories. As such we expect the AI Virtual Cell to contribute to a more unified view of biological processes, fostering alignment among scientists on how emergent properties in biology arise.

\looseness -1 By bridging the worlds of computer systems, modern generative AI and AI agents as well as biology, the AIVC could ultimately enable scientists to understand cells as information processing systems and build virtual depictions of life. As the AI Virtual Cell expands understanding of cellular and molecular systems, it will also increasingly allow us to program them and design novel synthetic ones. 
AI models have already been used to design new CRISPR enzymes \citep{ruffolo_2024}, functional proteins \citep{madani2023large}, and even entire prokaryotic genomes \citep{nguyen_2024}. The rapid progress in the precision of cell and genome engineering tools will accelerate this shift and different instantiations of the AIVC will compete in their ability to engineer new, functional biology capabilities as much as in their ability to represent and simulate biology \citep{khalil_2010}.

Finally, we staunchly advocate the role for open science approaches, where the scientific community readily shares data, models, and benchmarks, where findings and insights are contextualized, and where a climate of perpetual improvement is fostered. We welcome and encourage all stakeholders across sectors and domains to engage in this endeavor. With a massive scientific undertaking and shared goals, open sharing of insights and the power of safe, ethical and reliable AI, we believe we are stepping into a new era of scientific exploration and understanding. The confluence of AI and biology, as encapsulated by AI Virtual Cells, signals signals a paradigm shift in biology and shines as a beacon of optimism for unraveling multiple mysteries of the cell.

{\footnotesize
\bibliographystyle{naturemag}
\bibliography{references}
}

\newpage
\section*{Acknowledgements}
We thank Rok Sosi\v{c}, Charilaos Kanatsoulis, Lata Nair, Kexin Huang, Hanchen Wang, Minkai Xu, Michael Bereket, Romain Lopez, Takamasa Kudo, Ayush Agrawal, Arnuv Tandon, Mika Jain, Michihiro Yasunaga, Tim Jing, Michael Moor, George Crowley, Maria Brbić, Andrew Tolopko, Ivana Jelic, Ana-Maria Istrate, Sara Simmonds, Maximilian Lombardo, Pablo Garcia-Nieto, Mike Lin, Noorsher Ahmed, Orit Rozenblatt-Rosen, Gita Mahmoudabadi, Zoe Piran, Adam Gayoso and Anshul Kundaje for discussions. 
J.L. was supported by NSF under Nos. OAC-1835598 (CINES), CCF-1918940 (Expeditions), DMS-2327709 (IHBEM); Stanford Data Applications Initiative, Wu Tsai Neurosciences Institute, Stanford Institute for Human-Centered AI, Chan Zuckerberg Initiative, Amazon, Genentech, GSK, Hitachi, SAP, and UCB. E.L. was supported by Schmidt Futures, the Bridge2AI Program (NIH Common Fund; OT2 OD032742), the Cancer Cell Map Initiative (NCI Center for Cancer Systems Biology; U54 CA274502), the Wallenberg Foundation (2021.0346), Stanford Institute for Human-Centered AI, Chan Zuckerberg Initiative, and Param Hansa Philanthropies. 
\vfill\null
\columnbreak

\section*{Competing interests}
C.B. and A. R. are employees of Genentech, a member of the Roche Group. A.R. has equity in Roche. A.R. was a co-founder and equity holder of Celsius Therapeutics, and is an equity holder in Immunitas. Until July 31, 2020 A.R. was an S.A.B. member of ThermoFisher Scientific, Syros Pharmaceuticals, Neogene Therapeutics and Asimov. A.R. is a named inventor on multiple filed patents related to single cell and spatial genomics, including for scRNA-seq, spatial transcriptomics, Perturb-Seq, compressed experiments, and PerturbView. E.L. is an advisor for the Chan-Zuckerberg Initiative Foundation. N.J.S. is an employee of EvolutionaryScale, PBC.
\end{multicols}

\newpage
\begin{samepage}
{\newgeometry{margin=1in}
\begin{theo}[Grand challenges for building the AI Virtual Cell.]{box:challenges}
\paragraph{Outlining capabilities and designing evaluation frameworks.}
The burgeoning number of foundation models in biology perform a subset of the capabilities of virtual cells outlined in this Perspective. Given the diversity of these approaches, it is important to define what the core capabilities of AIVCs should be and how those capabilities can be evaluated. For every capability, proper metrics must be designed and comprehensive evaluation data be collected. Models' capabilities should be assessed both on general performance as well as on their ability to answer specific biological questions. It is imperative to continuously improve benchmarking strategies along with AIVC models and ensure that they align with biologically meaningful objectives. As the field develops better alignment on these questions, collaborative opportunities will arise and the speed at which virtual cells can be generated will accelerate.

\paragraph{Establishing self-consistency across varying contexts with different architectures.}
Biology is tremendously complex: it operates across different scales, in different contexts and is measured with different modalities. AIVC models must be self consistent across all of these axes. Models should propagate function across physical scales---interactions between molecules should have consistent effects when measuring binding affinity, gene expression, cell-cell communication or tissue organization. As physical and dynamic scales increase in scope and size, additional context, for example species, cell type, tissue, disease status etc, should fine-tune predictions made at smaller resolutions, while also accounting for stochasticity. Model predictions should also be agnostic to their input and output modalities. The same entity, profiled with different technologies, should have the same internal representation in an AIVC. To properly model such complex behaviors, many machine learning approaches should be explored and their merits carefully judged.



 

\paragraph{Balancing interpretability and  biological utility.}
A consistent trend in the application of deep learning methods to biology, accelerated by the rise of large foundation models, has been the implicit trade-off between models' performance gains and their increasingly uninterpretable 'black-box' natures. AIVC models will ultimately be judged on their ability to expand our understanding of biology, either by providing novel insights to biological processes or by accelerating the scientific process. To achieve this goal, AIVC models must make highly accurate and well-calibrated predictions that simulate biology, and the trade-off between actionabilty and interpretability will have to be balanced. Actionable model outputs are those of high utility to design affordable and efficient validation experiments and are key for initial real-world use. Various approaches exist for explaining model predictions, including causal modeling, sparse featurization and counterfactual reasoning, and this is a highly active research area. Building intuitive interfaces that facilitate the study and interpretation of AIVCs via other models, such as AI research agents will further increase downstream utility.



\paragraph{Constructing a framework for collaborative cell modeling.}
The successful development of AI Virtual Cells will require collaboration across disciplines. We foresee a future where AI Virtual Cell platforms function as open, interconnected hubs for collaborative development and broad deployment of cell models to researchers, and as education hubs delivering training to researchers, as well as providing engagement activities for educators, patients, and the public. Thus, investments in infrastructure fostering open and collaborative development of AI Virtual Cells should be of high priority. 

\paragraph{Ensuring AI Virtual Cells benefit all and promote ethical and responsible use.} 
Generating large open datasets that reflect human diversity---datasets integral for training AIVC models---poses a substantial challenge. Developers will have to use the utmost care to ensure these datasets are used ethically and transparently while building AIVCs and develop strategies to mitigate risks of model contamination with falsified data. Early adopters of AIVCs will have a key role in promoting and demonstrating responsible use of these models. Furthermore, the development of chat-based interfaces could be crucial in democratizing access to AIVCs. Close collaboration with ethics and regulatory experts from the outset is paramount for establishing new norms that will facilitate the responsible use of AI Virtual Cells.  

\paragraph{Understanding the value of different data types to prioritize large-scale data generation.}
A fundamental question for the collaborative development of AIVCs is which data and modalities should be collected to enable generalization across biological contexts and scales. These data will need to encompass the breadth of biology in different species, domains and modalities, representing the heterogeneity of life, while maintaining depth sufficient to distinguish true signals from noise. A key aspect of data generation will be the simultaneous measurement of temporal and physical scales, while also allowing perturbations of the system.

 \end{theo}
}
\clearpage

{\newgeometry{margin=1in}
\begin{theo}[Vignettes.]{box:vignettes}
\subsubsection*{Cell engineering to enable phenotypic drug discovery and cell-based therapeutics}
\begin{wrapfigure}{l}{0.45\textwidth}
\includegraphics[width=0.98\linewidth]{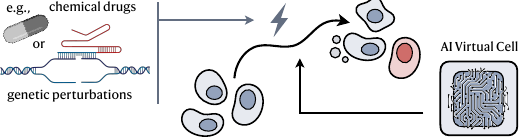} 
\end{wrapfigure}

One challenge in developing successful therapies is the difficulty in incorporating the full underlying genetic, molecular and cellular basis of disease during drug discovery and development \citep{nelson2015support}. These context-specific underpinnings are not fully specified, and often vary between human patients and model systems used in pre-clinical studies. By integrating biological data from various sources relevant to specific disease contexts, the AIVC could generate an environment for testing different therapeutic interventions \emph{in silico}, and identify approaches for engineering cells to reverse disease phenotypes, while accounting for the effects of varying both treatments and patient profiles. By representing the overall disease phenotype specific to patient populations (rather than one specific biochemical target at a time), the AIVC can enable virtual phenotypic screens. Even though \emph{in silico} experiments may not always be fully accurate, by prioritizing virtual hits with higher chances of success, the AIVC can lower experimentation costs and accelerate the process.

The AIVC has potential to push the cell therapy frontier. With growing evidence affirming the efficacy and safety of cell-based therapies for rare diseases and cancer \citep{mason2011cell, fischbach2013cell, bashor_2022}, the AIVC can improve systematization and precision to cell engineering. For example, virtual cell-based engineering could enable targeted modifications to pancreatic beta cells to create individualized beta cell replacement therapies for Type 1 diabetes. By simulating the biological phenotype of individual patients, \emph{in silico} experiments within the AIVC could identify interventions that help drive the differentiation of beta cells from progenitors, cloak them from the immune system, and maintain their function, with the ultimate goal of either transplanting these engineered cells into patients or engineering them \textit{in situ}.

\subsubsection*{Unlocking the power of spatial biology to fight cancer}
\begin{wrapfigure}{l}{0.45\textwidth}
\includegraphics[width=0.98\linewidth]{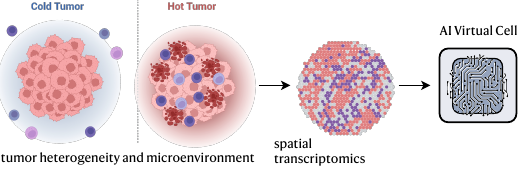} 
\end{wrapfigure}

Spatial structures in cancer, specifically within the tumor microenvironment (TME) are critical drivers of cancer progression, and can drive resistance to the immune system and limit drug efficacy \citep{jia_2022}. Malignant cells within a tumor can engage in active immune evasion, by either blocking immune infiltration \citep{melssen_2023}, evading immune recognition, or dampening immune cell function \citep{chow_2022}. Thus, immune resistance must be understood in the spatial context of the cellular neighborhood in order to identify the specific cell states and gene signatures involved. While next-generation spatial profiling methods enable researchers to experimentally investigate the heterogeneity of the TME \citep{de2023evolving}, an AIVC could extend these analyses to a universal, pan-cancer framework, which can be personalized to individual patients. Using an AIVC model, cancer researchers should be able to identify TME niches shared across multiple cancer types from many patients. Identifying pan-cancer markers can drive cancer treatment both by highlighting new targets, but also by identifying existing treatments that can be applied to new cancer types \citep{kinker_2020, barkley_2022}. 
In this setting, the AIVC would help identify the interactions associated with TME cell states, and would search for similar states from any disease where existing treatments exist.

Finally, the AIVC could greatly advance precision oncology \citep{schwartzberg2017precision}. Given that the AIVC will capture intrinsic variation, the genetic diversity of individual patient's cancers will be represented in any analyses. While the AIVC would already accurately qualify the change in the expression of genes, tumor sequencing data would allow it to model the change of function of those genes, for example through loss of function, change in post-translational modifications, or rewiring of protein-protein interactions and signaling networks \citep{aebersold2018many}.

\subsubsection*{Diagnostic virtual cell models for individual patients}
\begin{wrapfigure}{l}{0.45\textwidth}
\includegraphics[width=0.98\linewidth]{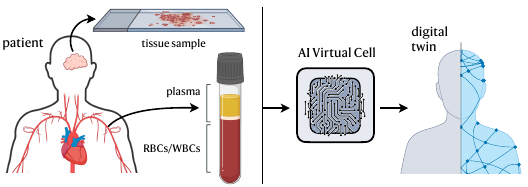} 
\end{wrapfigure}

The AIVC could introduce a new approach to diagnostics that incorporates a personalized AI Virtual Cell (or a digital twin \citep{katsoulakis2024digital}) to track a patient's health and suggest suitable interventions. The AIVC would create a detailed representation of each patient's cells by incorporating specific patient data, such as genetic sequences, single-cell profiles from blood, and tissue pathology images, along with additional clinical information from their health records. Periodic updates to each patient's AIVC instance enable monitoring of evolving health conditions, prediction of upcoming adverse events and potential therapeutic outcomes. 

Through additional updates from less costly assays, this virtual patient model could be progressively refined and made more robust \citep{rajewsky2020lifetime}. For example, transcriptomic or genetic liquid biopsies can reveal significant and diverse characteristics of a patient from a single test and could greatly aid in the diagnosis of a broad spectrum of conditions \citep{alix2021liquid}. Through the virtual cell's implicit and structured representation of universal cell types and states, one can envision the creation of patient models of inaccessible cell types, such as beta cells in the pancreas or neurons in the brain, generated after sampling accessible cell types such as blood or skin.

\subsubsection*{A hypothesis-generating framework for scientific research}
\begin{wrapfigure}{l}{0.45\textwidth}
\includegraphics[width=0.98\linewidth]{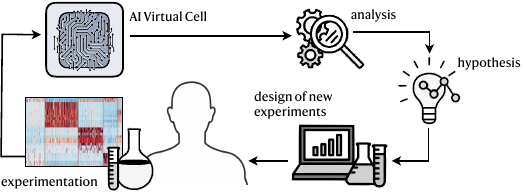} 
\end{wrapfigure}

Traditionally, the biological research community has relied on computational models for analyzing data from past experiments based on an existing hypothesis \citep{raue2013lessons, covert2004integrating}. The virtual cell could switch the paradigm by computationally exploring a vast array of possible hypotheses through \emph{in silico} experimentation. It could identify the most informative experiments for addressing specific biological questions, shifting the role of computational models from merely validating hypotheses or processing observations without a particular goal to generating specific sets of hypotheses to pursue.

This shift could greatly enhance the scientific discovery process: Instead of conducting a single experiment followed by an in-depth analysis, scientists can engage in a dynamic iterative interaction with the virtual cell. With each new piece of data, they can refine their understanding of the biological system and consult the virtual cell to identify what additional experimental data could be valuable. Ultimately, we may be able to perform active learning with biologists in the loop and construct self-driving labs for efficient and unbiased generation of virtual cells. 
\end{theo}
}

{\newgeometry{margin=1in}
\begin{theo}[AI techniques for building the AI Virtual Cell.]{box:architectures}
\looseness -1 The AI Virtual Cell will connect a number of diverse neural network architectures. While these architectures may not have been purpose-built for biological applications, they have each demonstrated success when matched with specific biological modalities and inductive biases. In many cases, these architectures may be exchangeable, and one must weigh their individual trade-offs in accuracy, speed, and generalizability. Beyond, the community is actively developing AI architectures tailored to the characteristics of (large) biological datasets.

\subsubsection*{Transformers}
\begin{wrapfigure}{l}{0.33\textwidth}
\includegraphics[width=0.98\linewidth]{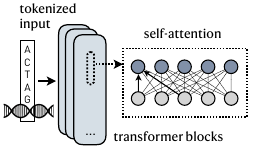} 
\end{wrapfigure}
A transformer neural network \citep{vaswani_2017} is comprised of multiple transformer layers, each taking a series of tokens (discrete pieces of information such as words, RNA molecules, or gene representations) as input---initial tokens for the first layer and outputs from the preceding layer for subsequent ones. Within each layer, tokens use self-attention to integrate context from other tokens, enhancing their own representations, which are then processed through a feed forward network. This architecture, which fundamentally requires only a collection of tokens, adapts well across various applications and use cases.

The collection of tokens passed to a transformer does not have any ordering by default. Additionally, the self attention mechanism, the core of the success of the transformer, can be taken as a strong biological inductive bias. For instance, in representing cells through their RNA molecules detected via scRNA-seq, each RNA molecule, represented as a token, interacts with others, modeling gene interactions through self-attention \citep{vaswani_2017}. Customizing input tokens with numerical representations of genes further allows the integration of diverse biological data scales, from individual genes to whole cells \citep{rosen_2023, cui2024scgpt, chen_2024_genept}.

Additionally, introducing positional encodings to tokens enables transformers to process sequences, such as natural language \citep{vaswani_2017} or biological sequences like DNA \citep{ji_2021, dallatorre_2023, vaishnav2022evolution}, by incorporating sequence-specific dependencies. This approach is crucial in applications like masked language modeling, where the model predicts missing tokens in sequences, enhancing its understanding of contextual relationships within data. Innovations continue to refine transformers, increasing their capacity to handle longer sequences and improving efficiency, with advancements like state-space models enabling the generation of extensive DNA sequences \citep{nguyen_2023, nguyen_2024}. 

\subsubsection*{Convolutional neural networks}
\begin{wrapfigure}{l}{0.33\textwidth}
\includegraphics[width=0.98\linewidth]{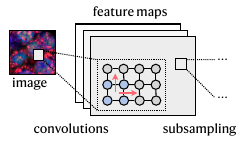} 
\end{wrapfigure}
A convolutional neural network is a deep learning model primarily used for analyzing images \citep{fukushima_1980, lecun_1995}. It consists of multiple layers that automatically and adaptively learn spatial hierarchies of features through backpropagation. This learning is facilitated by convolutional layers that apply filters to local patches of input data, pooling layers that reduce dimensionality, and fully connected layers that interpret the features extracted to make decisions.

In the field of biology, CNNs have proven invaluable for tasks involving image data due to their ability to detect complex patterns and structures, such as microscope images of cells and tissues. Here, CNNs play a critical role in multiplex imaging \citep{gomez2021deepimagej}, where multiple targets within a single sample are labeled and visualized simultaneously. This technique is particularly useful in studying the complex interactions of different molecules or cell types within a heterogeneous tissue environment \citep{le2022analysis}. Another notable application is in the analysis of H\&E stained tissue sections, commonly used in clinical pathology \citep{chen2024towards}. Lastly, in live-cell imaging, CNNs are employed to track dynamic changes within cells or even single-molecules over time, providing insights into cell migration, responses to treatment or the movement and interaction of individual molecules within cells, revealing crucial biological processes at a molecular level \citep{moen2019deep}. 

Beyond their traditional use in image processing, CNNs can also be applied to model sequence data, such as DNA sequences, where they identify patterns and features that are predictive of biological functions \citep{avsec_2021}.
Despite their extensive utility, CNNs are increasingly being supplemented or replaced by vision transformer models \citep{dosovitskiy_2020}, which leverage self-attention mechanisms to process entire images in parallel. These models can often achieve higher accuracy on tasks where understanding the global context within the image is crucial.

\newpage
\subsubsection*{Diffusion models}
\begin{wrapfigure}{l}{0.32\textwidth}
\includegraphics[width=0.98\linewidth]{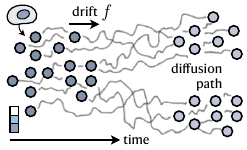} 
\end{wrapfigure}
Diffusion models are a class of generative deep learning models that have recently gained attention for their ability to generate high-quality, diverse samples across various domains \citep{ho_2020}. They operate by gradually transforming a distribution of random noise into a structured output (e.g., images, text, cellular states, etc.) through a process that mimics a physical diffusion process.
Building up on diffusion model architectures, approaches such as flow matching methods can also model the distributional evolution over time \citep{lipman2023flow}, making them especially powerful in biological applications where dynamic changes and temporal progression are critical. Flow matching methods thus capture and generate sequences of data that reflect continuous transformations, such as the developmental stages of cells over time and space or the response of biological systems to treatments \citep{somnath2023aligned, pariset2023unbalanced}.
The ability of diffusion and flow matching models to learn and replicate complex distributions, combined with the temporal and spatial modeling capabilities of flow matching methods, makes them particularly suited for tasks that involve high-dimensional, intricate data structures typical of biological systems.

\subsubsection*{Graph neural networks}
\begin{wrapfigure}{l}{0.33\textwidth}
\includegraphics[width=0.98\linewidth]{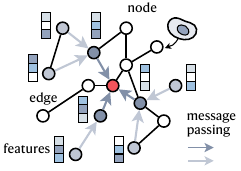} 
\end{wrapfigure}
\looseness-1 Graph neural networks are a set of architectures that can model graphical data \citep{scarselli_2009}. Graphs, sets of nodes connected by edges, are useful representations for many kinds of biological data. When modeling a biological system, a GNN could be a good choice if a graph structure represents some core inductive bias. For example, a protein structure \citep{cao_2020} can be thought of as a graph where residues are nodes, and their bonds are edges. Cells in a tissue form a graph: each cell is a node, and the cells it is physically proximal to, are connected by edges \citep{brbi_2022, wu_2022}. In both cases, the graph represents how nodes are physically proximal to each other. For spatially organized cells, the graph represents how they may pass chemical signals between one another. 

\looseness -1 GNNs can be used to make predictions about individual nodes, edges or the graph as a whole \citep{hamilton2017inductive}. For simplicity, in the following, we describe a node-based GNN. At each layer, a node updates its representation using a neural network, which can take in that node's current representation, in addition to the representations of the node's neighbors, which are connected by an edge. By stacking GNN layers, a node can receive 'messages` from neighboring nodes at increasing distances, 'hops`, from it. Nodes and edges can both be initialized with different features, which control their final representation and what messages they pass to their neighbors. For example, a GNN trained on spatial transcriptomic data could take node features to be the virtual cell representation of each cell's gene expression. The GNN would then update those representations to include context about each cell's neighbors, helping to identify spatial interactions and niches \citep{wu_2022}.
\end{theo}
}
\end{samepage}

\end{document}